\documentclass[11pt]{article}
\usepackage[top=1.0in, bottom=1.0in, left=1.0in, right=1.0in]{geometry}
\usepackage{tikz}
\usepackage{braket}
\usepackage{amsmath}
\usepackage{algorithm}
\usepackage[noend]{algpseudocode}
\usepackage{amssymb}
\usepackage{multicol}
\usetikzlibrary{arrows, shapes, intersections, backgrounds, positioning, automata, calc, trees}
\usepackage{graphicx}
\usepackage{parskip}
\usepackage{titling}
\usepackage{qcircuit}
\usepackage{sectsty}
\usepackage{color}
\usepackage{listings}
\usepackage[colorinlistoftodos]{todonotes}
\usepackage{colortbl}
\usepackage{caption}

\usepackage{fancyvrb}
\makeatletter
\def\BState{\State\hskip-\ALG@thistlm}
\makeatother

\title{A Multilayer Network Approach to Quantum Computing}
\author{Perry Sakkaris, QuDot Inc. \\ Ryan Sudhakaran, QuDot Inc.}
\date{\vspace{-5ex}}

\begin{document}
	\newcommand*{\rom}[1]{\expandafter\@slowromancap\romannumeral #1@}
		\tikzset{vertex/.style = {shape=circle,draw,minimum size=2.0em}}
	\tikzset{root vertex/.style = {vertex, fill=black!50}}
	\tikzset{edge/.style = {->,> = latex',thick}}
	\tikzset{cross/.style = {pos=0.1}}
	\tikzset{line/.style={draw,thick,-latex', shorten >=2pt}}
	\tikzset{invisible/.style={minimum width=0mm,inner sep=0mm,outer sep=0mm}}
	
	\tikzset{me/.style={to path={
				\pgfextra{% 
					\pgfmathsetmacro{\startf}{-(#1-1)/2}  
					\pgfmathsetmacro{\endf}{-\startf} 
					\pgfmathsetmacro{\stepf}{\startf+1}}
				\ifnum 1=#1 -- (\tikztotarget)  \else
				let \p{mid}=($(\tikztostart)!0.5!(\tikztotarget)$) 
				in
				\foreach \i in {\startf,\stepf,...,\endf}
				{%
					(\tikztostart) .. controls ($ (\p{mid})!\i*6pt!90:(\tikztotarget) $) .. (\tikztotarget)
				}
				\fi   
				\tikztonodes
	}}}

	\maketitle
	\begin{abstract}
		The circuit model of quantum computation is reformulated as a multilayer network theory \cite{boccaletti2014structure} called a Quantum Multiverse Network (QuMvN). The QuMvN formulation allows us to interpret the quantum wave function as a combination of ergodic Markov Chains where each Markov Chain occupies a different layer in the QuMvN structure. Layers of a QuMvN are separable components of the corresponding wave function. Single qubit measurement is defined as a state transition of the Markov Chain that emits either a $0$ or $1$ making each layer of the QuMvN a Discrete Information Source. A message is equivalent to a possible measurement outcome and the message length is the number of qubits. Therefore, the quantum wave function can be treated as a combination of multiple discrete information sources analogous to what Shannon called a ``mixed'' information source \cite{shannon1948mathematical}.  We show the QuMvN model has significant advantages in the classical simulation of some quantum circuits by implementing quantum gates as edge transformations on the QuMvNs. We implement a quantum virtual machine capable of simulating quantum circuits using the QuMvN model and use our implementation to classically simulate Shor's Algorithm \cite{shor1999polynomial}. We present results from multiple simulations of Shor's Algorithm culminating in a $70$ qubit simulation of Shor's Algorithm on a commodity cloud server with $96$ CPUS and $624$GB of RAM. Lastly, the source of quantum speedups is discussed in the context of layers in the QuMvN framework and how randomized algorithms can push the quantum supremacy boundary.
		
	\end{abstract}	
	\section{Introduction}
	We define Quantum Multiverse Networks (QuMvNs) that formulate quantum computers in the mathematical language of Multilayer Network Theory. The discipline of network science has shown us that important information of complex systems can be obtained by studying the structure and function of an underlying multilayer network model \cite{boccaletti2014structure}. Also, we have seen ideas from quantum physics being used in computational aspects of complex networks \cite{de2015structural}. Significant progress has been made recently in applying complex network models in quantum information \cite{biamonte2019complex}. QuMvNs are such a model for quantum computation. 
	Degrees of freedom for a quantum system can be modeled with nodes, edges and layers in a type of multilayer network called a multiplex network \cite{boccaletti2014structure}. QuMvNs describe qubits by a fixed set of graphical nodes labeled $0$ or $1$ representing degrees of freedom for a single qubit. An edge in the network represents a tensor product between qubits and weights encode the probability amplitude to observe the child node given the parent. We show that each network layer satisfies the Markov Property and thus is a Markov Chain. 
	
	If the Markov Chain is kept ergodic and emits a letter in the alphabet $\{0,1\}$ for every transition, then we satisfy Shannon's criterion of a ``discrete information source'' \cite{shannon1948mathematical}. First, we keep Markov Chains ergodic by using a different layer of the QuMvN for entangled states. Second, we implement single qubit measurement by walking the Markov Chain using the Independent Cascade Information Diffusion Model (ICM) \cite{gruhl2004information}. As we visit a node we emit the node value $\{0,1\}$. Having ergodic Markov Chains in each layer of a QuMvN with the ICM as a measurement model corresponds to Shannon's description of multiple discrete information sources where we first choose an information source and then produce a message from the source \cite{shannon1948mathematical}.

	A QuMvN can implement single qubit gates by transforming edge weights (probability amplitudes) of network edges. We demonstrate how to implement $X, H$ and $R(k)$ gates. We then show how control gates spawn new layers in our QuMvN model. We examine how the growth rate in the number of layers $L$ in a QuMvN is the key property that makes the classical simulation of quantum circuits difficult. If $O(L)$ can be kept polynomial in the number of qubits $n$ through randomization algorithms then it is possible to simulate quantum computation on classical hardware beyond the capabilities of current simulation methods. We found that our QuMvN formalism is well suited for the classical simulation of Shor's Algorithm \cite{shor1999polynomial}. We present multiple results from $35$-$70$ qubit simulations of Shor's Algorithm using our QuMvN framework. To our knowledge, $70$ qubits is the largest simulation of Shor's Algorithm to date. Lastly, we discuss how this new point of view can help shed light on the ``elusive source of quantum speedups'' \cite{vedral2010elusive}.

	\section{Single Layer QuMvN}
	We begin by developing our methods for a single layer qubit QuMvN in the computational basis. A single layer QuMvN is able to represent a separable qubit wave function $\ket{\Psi}$ that admits a tensor factorization $\ket{\Psi} = \ket{\psi_1} \otimes \ket{\psi_2} \dotsb \otimes \ket{\psi_n}$ where $n$ is the number of qubits. Therefore, a single layer QuMvN can represent a non-entangled state. We restrict our analysis to qubit systems in the computational basis $\{\ket{0}, \ket{1}\}$ and assume $\ket{\Psi}$ can be obtained by starting in the initial state $\ket{0}^{\otimes n}$ and applying a quantum circuit.
	
	\subsection{Representation}
	A single layer of a QuMvN for qubit systems can be represented by a directed acyclic graph $Q=\{V, E\}$ where $V, E$ are the set of vertices and edges respectively. The vertices $V$ represent the degrees of freedom of the system and are labeled by an alphabet $A$. Therefore, for qubits we have two vertices labeled from the alphabet $A=\{0,1\}$ per qubit. The edges have a weight $\in \mathbb{C}$ and represent the probability amplitude of the destination vertex given the source vertex. 
	 
	 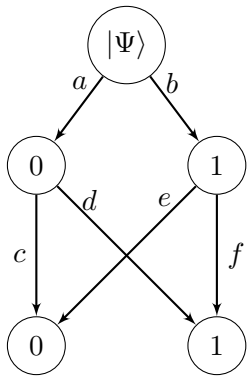
\begin{figure}[!h]
	 	\centering
	 	\begin{tikzpicture}[scale=0.8]
	 	\node[vertex] (r) at  (1.5,2) {$\ket{\Psi}$};
	 	
	 	\node[vertex] (00) at  (0,0) {$0$};
	 	\node[vertex] (01) at  (3,0) {$1$};	
	 	\node[vertex] (10) at  (0,-3) {$0$};
	 	\node[vertex] (11) at  (3,-3) {$1$};
	 	
	 	\draw[edge, cross] (r) edge node[left]{$a$} (00);
	 	\draw[edge, cross] (r) edge node[right]{$b$} (01);
	 	\draw[edge] (00) edge node[left]{$c$}  (10);
	 	\draw[edge,cross] (01) edge node[left]{$e$}  (10);
	 	\draw[edge,cross] (00) edge node[right]{$d$}  (11);
	 	\draw[edge] (01) edge node[right]{$f$}  (11);
	 	\end{tikzpicture}
	 	\caption{ $\ket{\Psi}=ac\ket{00} + ad\ket{01} + be\ket{10} + bf\ket{11}$}
	 	\label{fig:f77}
	 \end{figure}	
 
 	  We begin with the generic two qubit state in FIG.~\ref{fig:f77} where a possible measurement outcome is  equivalent to a possible path traversal of the single layer network and the probability amplitude of that traversal is a product of the edge weights on the path traversed. The separability condition of single layer QuMvNs places restrictions on the possible values of the edge weights. For $\ket{\Psi}=ac\ket{00} + ad\ket{01} + be\ket{10} + bf\ket{11}$ separability implies $c=e$, $d=f$ and $\ket{\Psi}=\left(a\ket{0} + b\ket{1}\right) \otimes \left(c\ket{0} + d\ket{1}\right)$. In general, the separability criterion of single layer QuMvNs require equality of edge weights pointing to the same vertex. The probability of obtaining a state is the magnitude square of the probability amplitude. For example, $P\left(\ket{00}\right) = \left|ac\right|^2 = \left|a\right|^2 \times \left|c\right|^2.$
 
	  We illustrate single layer QuMvNs by developing a few examples. For simplicity, edges with an edge weight of $0$ are omitted. FIG.~\ref{fig:f1} shows the one qubit system $\ket{\Psi} = \frac{1}{\sqrt{2}} \left( \ket{0} + \ket{1}\right)$. Figure ~\ref{fig:f2} shows the two qubit state $\ket{\Psi}=\frac{1}{\sqrt{2}}\ket{00} + \frac{1}{\sqrt{2}}\ket{10}$, and FIG. ~\ref{fig:f3} shows the state  $\ket{\Psi}=\frac{1}{2} \left(\ket{00} + \ket{01} + \ket{10} + \ket{11}\right)$. Lastly, we show the $n$ qubit superposition $\ket{\Psi}=H^{\otimes n} \ket{0}^{\otimes n}$ (where $H$ is the Hadamard Gate) in FIG. ~\ref{fig:f4}.
			
	\begin{figure}[!h]
		\centering
		\begin{tikzpicture}[scale=0.8]
		\node[vertex] (r) at  (1.5,2) {$\ket{\Psi}$};

		\node[vertex] (00) at  (0,0) {$0$};
		\node[vertex] (01) at  (3,0) {$1$};	
		
		\draw[edge, cross] (r) edge node[left]{$\frac{1}{\sqrt{2}}$} (00);
		\draw[edge, cross] (r) edge node[right]{$\frac{1}{\sqrt{2}}$} (01);
		\end{tikzpicture}
		\caption{QuMvN representation of $\ket{\Psi} = \frac{1}{\sqrt{2}} \left( \ket{0} + \ket{1}\right)$}
		\label{fig:f1}
	\end{figure}
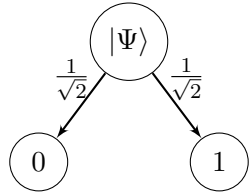

\begin{figure}[!h]
	\centering
	\begin{tikzpicture}[scale=0.8]
	\node[vertex] (r) at  (1.5,2) {$\ket{\Psi}$};

	\node[vertex] (00) at  (0,0) {$0$};
	\node[vertex] (01) at  (3,0) {$1$};	
	\node[vertex] (10) at  (0,-3) {$0$};
	\node[vertex] (11) at  (3,-3) {$1$};
	
	\draw[edge, cross] (r) edge node[left]{$\frac{1}{\sqrt{2}}$} (00);
	\draw[edge, cross] (r) edge node[right]{$\frac{1}{\sqrt{2}}$} (01);
	\draw[edge] (00) edge node[left]{$1$}  (10);
	\draw[edge,cross] (01) edge node[left]{$1$}  (10);
	\end{tikzpicture}
	\caption{The QuMvN encodes the factorization of $\ket{\Psi}=\frac{1}{\sqrt{2}}\ket{00} + \frac{1}{\sqrt{2}}\ket{10}=\left(\frac{1}{\sqrt{2}}\ket{0} + \frac{1}{\sqrt{2}}\ket{1}\right) \otimes \ket{0}$}
	\label{fig:f2}
\end{figure}
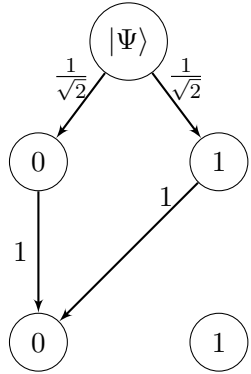

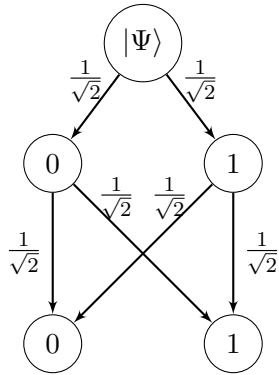
\begin{figure}[!h]
	\centering
	\begin{tikzpicture}[scale=0.8]
	\node[vertex] (r) at  (1.5,2) {$\ket{\Psi}$};
	
	\node[vertex] (00) at  (0,0) {$0$};
	\node[vertex] (01) at  (3,0) {$1$};	
	\node[vertex] (10) at  (0,-3) {$0$};
	\node[vertex] (11) at  (3,-3) {$1$};
	
	\draw[edge, cross] (r) edge node[left]{$\frac{1}{\sqrt{2}}$} (00);
	\draw[edge, cross] (r) edge node[right]{$\frac{1}{\sqrt{2}}$} (01);
	\draw[edge] (00) edge node[left]{$\frac{1}{\sqrt{2}}$}  (10);
	\draw[edge,cross] (01) edge node[left]{$\frac{1}{\sqrt{2}}$}  (10);
	\draw[edge,cross] (00) edge node[right]{$\frac{1}{\sqrt{2}}$}  (11);
	\draw[edge] (01) edge node[right]{$\frac{1}{\sqrt{2}}$}  (11);
	\end{tikzpicture}
	\caption{ $\ket{\Psi}=\frac{1}{2} \left(\ket{00} + \ket{01} + \ket{10} + \ket{11}\right)=\left(\frac{1}{\sqrt{2}}\ket{0} + \frac{1}{\sqrt{2}}\ket{1}\right) \otimes \left(\frac{1}{\sqrt{2}}\ket{0} + \frac{1}{\sqrt{2}}\ket{1}\right)$}
	\label{fig:f3}
\end{figure}	 

	\begin{figure}[!h]
		\centering
		\begin{tikzpicture}[scale=0.8]
		\node[vertex] (r) at  (1.5,2) {$\ket{\Psi}$};
		
		\node[vertex] (00) at  (0,0) {$0$};
		\node[vertex] (01) at  (3,0) {$1$};	
		
		\node[vertex] (10) at  (0,-3) {$0$};
		\node[vertex] (11) at  (3,-3) {$1$};	
		
		\node[vertex] (20) at  (0,-6) {$0$};
		\node[vertex] (21) at  (3,-6) {$1$};	
		
		\node[vertex] (n0) at  (0,-9) {$0$};
		\node[vertex] (n1) at  (3,-9) {$1$};
		
		\draw[edge, cross] (r) edge node[left]{$\frac{1}{\sqrt{2}}$} (00);
		\draw[edge, cross] (r) edge node[right]{$\frac{1}{\sqrt{2}}$} (01);
		
		\draw[edge] (00) edge node[left]{$\frac{1}{\sqrt{2}}$}  (10);
		\draw[edge,cross] (00) edge node[right]{$\frac{1}{\sqrt{2}}$}  (11);
		
		\draw[edge, cross] (01) edge node[left]{$\frac{1}{\sqrt{2}}$} (10);
		\draw[edge] (01) edge node[right]{$\frac{1}{\sqrt{2}}$} (11);
		
		\draw[edge] (10) edge node[left]{$\frac{1}{\sqrt{2}}$} (20);
		\draw[edge, cross] (10) edge node[right]{$\frac{1}{\sqrt{2}}$} (21);
		
		\draw[edge, cross] (11) edge node[left]{$\frac{1}{\sqrt{2}}$} (20);
		\draw[edge] (11) edge node[right]{$\frac{1}{\sqrt{2}}$} (21);
		
		\draw[line, black, densely dashed] (-1,-1) rectangle (4,1);
		\node (q1) at (-2,0) {$qubit\phantom{a}1$};
		
		\draw[line, black, densely dashed] (-1,-4) rectangle (4,-2);
		\node (q2) at (-2,-3) {$qubit\phantom{a}2$};
		
		\draw[line, black, densely dashed] (-1,-7) rectangle (4,-5);
		\node (q3) at (-2,-6) {$qubit\phantom{a}3$};
		
		\draw[line, black, densely dashed] (-1,-10) rectangle (4,-8);
		\node (q4) at (-2,-9) {$qubit\phantom{a}n$};
		
		\path (20) to node {\vdots} (n0);
		\path (21) to node {\vdots} (n1);
		\end{tikzpicture}
		\caption{$\ket{\Psi}=\left(\frac{1}{\sqrt{2}}\ket{0} + \frac{1}{\sqrt{2}}\ket{1}\right) \otimes \left(\frac{1}{\sqrt{2}}\ket{0} + \frac{1}{\sqrt{2}}\ket{1}\right) \otimes \dotsb \otimes
			\left(\frac{1}{\sqrt{2}}\ket{0} + \frac{1}{\sqrt{2}}\ket{1}\right)$}
		\label{fig:f4}
	\end{figure}
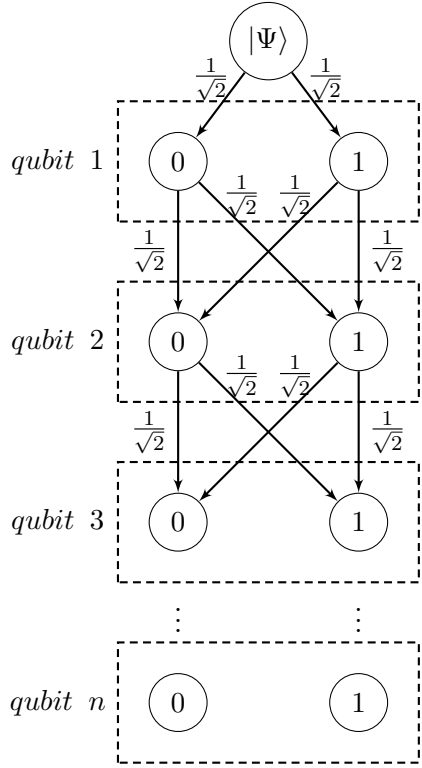

The examples in FIG. 2-5 show the advantage of representing separable quantum states as QuMvNs instead of vectors in Hilbert Space $\mathcal{H}$. Given an $n$ qubit separable state the vector representation in $\mathcal{H}$ requires $2^n$ elements whereas the QuMvN representation requires only $2n+1$ nodes and $4n-2$ edges with complex weights. The complex weights play an important role in QuMvNs as they are the probability amplitude of the destination qubit \textit{value} given the source qubit \textit{value}.

Suppose we are given the wave function in FIG. ~\ref{fig:f3}: $\ket{\Psi}=\frac{1}{2} \left(\ket{00} + \ket{01} + \ket{10} + \ket{11}\right)$.
Our initial condition is $P\left(\ket{\Psi} \right) = 1$. The probability of qubit one being $\ket{0}$, $P\left( q_1=\ket{0} | \Psi\right)$, is given by the probability amplitude connecting root node $\ket{\Psi}$ with the node labeled $0$ of $q_1$. Therefore, $P\left( q_1=\ket{0} | \Psi\right) = \left| \frac{1}{\sqrt{2}}\right|^2 = \frac{1}{2}$. 

 We can calculate probabilities of the possible values of $q_2$ given $\Psi$ with 
\begin{align*}
P\left(q_2=\ket{0} | \Psi\right) &= P\left(q_2=\ket{0} | q_1=\ket{0}\right)P\left(q_1=\ket{0}|\Psi\right) + P\left(q_2=\ket{0} | q_1=\ket{1}\right)P\left(q_1=\ket{1}|\Psi\right) \\
  &= \left(\left|\frac{1}{\sqrt{2}}\right|^2 \times \left|\frac{1}{\sqrt{2}}\right|^2\right) + \left(\left|\frac{1}{\sqrt{2}}\right|^2 \times \left|\frac{1}{\sqrt{2}}\right|^2\right)\\
                          &= \frac{1}{4} + \frac{1}{4} \\
                          &= \frac{1}{2}
\end{align*}

We can calculate the probability of a possible result for an $n$ qubit state by walking the QuMvN and chaining the probability amplitudes of the edges on a path. The probability of an $n$ qubit result is the product of $n$ conditional probabilities corresponding to $n$ single qubit measurements. An overview of chaining probability amplitudes for a sequence of measurements $a,b,c,d...k$ with $P_{a,b,c,d..k}=\left|\phi_{abcd...k} \right|^2$ is given by Feynman in \cite{feynman2005space}. 

Let $\ket{m_1m_2...m_n}$ be a possible state of an $n$ qubit wave function $\ket{\Psi_n}$ where $m_1,m_2,...,m_n \in \{0,1\}$. Then
\begin{align}
%\begin{equation}
P\left(\ket{m_1m_2...m_n}\right)&=
P\left(q_1=\ket{m_1}|\ket{\Psi_n}\right) \nonumber \\ 
&\times P\left(q_2=\ket{m_2}|q_1=\ket{m_1}\right) \nonumber\\
& \cdots \nonumber\\ 
&\times P\left(q_n=\ket{m_n}|q_{n-1}=\ket{m_{n-1}}\right)
%\end{equation}
\end{align}

Conditional probabilities can be calculated as shown in Equation 1.  Note that Equation 1 is the familiar probability rule for multiple measurements described by Feynman, also note the similarity of Feynman's description and the product rule of Bayesian Networks \cite{koller2009probabilistic}.

To illustrate, the probability of observing $\ket{10}$ given by $\ket{\Psi}$ in example 3 is:
\begin{align*}
		P\left(\ket{10}|\Psi\right) &= P\left(q_1=\ket{1} | \Psi\right)P(q_2=\ket{0} | q_1 = \ket{1}) = \frac{1}{4} 
\end{align*}

Using these results for calculating conditional probabilities we are able to define the measurement process on a single layer QuMvN.

	\subsection{Measurement}
	A single measurement result of an $n$ qubit QuMvN representing $\ket{\Psi}$ is obtained by $n$ single-qubit measurements starting from vertex $\ket{\Psi}$ and ending at qubit $n$ using the Independent Cascade Information Diffusion Model (ICM) \cite{gruhl2004information} (a type of random walk). At each step of the walk we emit the destination vertex label. Starting from vertex $\ket{\Psi}$ we are given the initial condition of the walk to be $P\left(\ket{\Psi}\right)=1$. We take one step to a neighbor vertex at random according to the probability amplitudes of edges joining the current vertex to its neighbors. For $m,j \in \{0,1\}$, from equation $(1)$ we have the probability of measuring the next qubit, $P\left(q_{i+1}=\ket{j}\right)$, is dependent on the value of the current qubit $q_i=\ket{m}$. Let:
	
	\begin{equation}
		p\left(q_{i+1}, m, j\right) = P\left(q_{i+1}=\ket{j} | q_{i}=\ket{m}\right)
	\end{equation}
	
	where the first qubit $p\left(q_1, m, j\right) = p\left(q_1, m\right) = P\left(q_1=\ket{m} | \Psi\right)$

	\begin{algorithm}
		\floatname{algorithm}{Method}
		\caption{QuMvN Single Layer Measurement}
		\begin{algorithmic}[1]
			\Procedure{measure state}{}
				\State{$m \gets 0$}
				\State{$bitstring \gets empty$}
				\For{$q_i \gets 1$ to $numQubits$}
				    \State{$rand \gets$ random number between $[0,1]$ }
				    \If{$rand <= p\left(q_i, m, 0\right)$}
				        \State{$m \gets 0$}
				    \Else
				    	\State{$m \gets 1$}
				    \EndIf
				    \State{$bitstring \gets concat\left(bitstring, m\right)$}
				\EndFor
				
				\Return{$bitstring$}
			\EndProcedure
		\end{algorithmic}
	\end{algorithm}

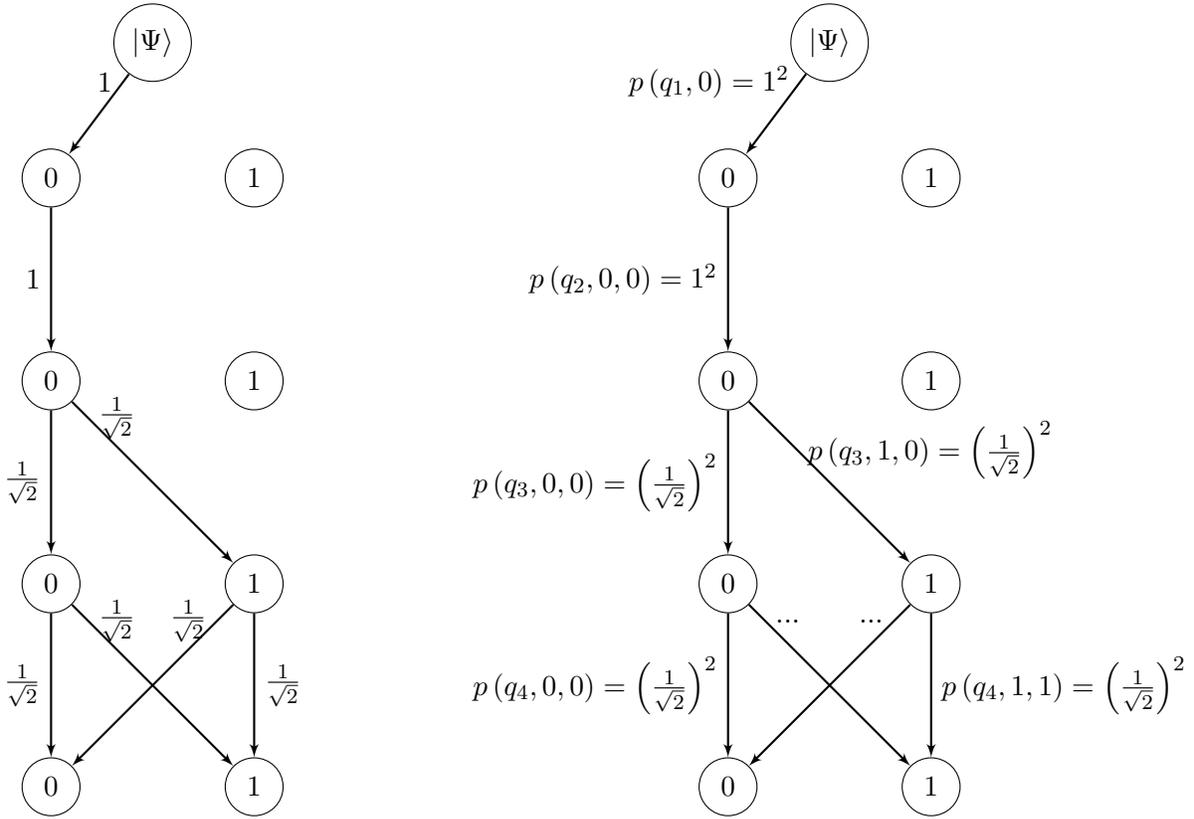
\begin{figure}[!h]
    	\centering
    	\begin{tikzpicture}[scale=0.9]
    	\node[vertex] (r) at  (1.5,2) {$\ket{\Psi}$};
    	
    	\node[vertex] (00) at  (0,0) {$0$};
    	\node[vertex] (01) at  (3,0) {$1$};	
    	\node[vertex] (10) at  (0,-3) {$0$};
    	\node[vertex] (11) at  (3,-3) {$1$};
    	\node[vertex] (20) at  (0,-6) {$0$};
    	\node[vertex] (21) at  (3, -6) {$1$};	
    	\node[vertex] (30) at  (0,-9) {$0$};
    	\node[vertex] (31) at  (3,-9) {$1$};
    	
    	\draw[edge, cross] (r) edge node[left]{$1$} (00);
    	\draw[edge] (00) edge node[left]{$1$}  (10);
    	\draw[edge,cross] (10) edge node[right]{$\frac{1}{\sqrt{2}}$}  (21);
    	\draw[edge] (10) edge node[left]{$\frac{1}{\sqrt{2}}$}  (20);
    	\draw[edge] (20) edge node[left]{$\frac{1}{\sqrt{2}}$}  (30);
    	\draw[edge,cross] (20) edge node[right]{$\frac{1}{\sqrt{2}}$}  (31);
    	\draw[edge,cross] (21) edge node[left]{$\frac{1}{\sqrt{2}}$}  (30);
    	\draw[edge] (21) edge node[right]{$\frac{1}{\sqrt{2}}$}  (31);

    	\node[vertex] (r2) at  (11.5,2) {$\ket{\Psi}$};

    	\node[vertex] (200) at  (10,0) {$0$};
    	\node[vertex] (201) at  (13,0) {$1$};	
    	\node[vertex] (210) at  (10,-3) {$0$};
    	\node[vertex] (211) at  (13,-3) {$1$};
    	\node[vertex] (220) at  (10,-6) {$0$};
    	\node[vertex] (221) at  (13, -6) {$1$};	
    	\node[vertex] (230) at  (10,-9) {$0$};
    	\node[vertex] (231) at  (13,-9) {$1$};
    	
    	\draw[edge, cross] (r2) edge node[left]{$p\left(q_1, 0\right) = 1^2$} (200);
    	\draw[edge] (200) edge node[left]{$p\left(q_2,0,0\right)=1^2$}  (210);
    	\draw[edge,cross] (210) edge node[right, style={pos=0.3}]{$p\left(q_3, 1, 0\right) = \left(\frac{1}{\sqrt{2}}\right)^2$}  (221);
    	\draw[edge] (210) edge node[left]{$p\left(q_3, 0, 0\right) = \left(\frac{1}{\sqrt{2}}\right)^2$}  (220);
    	\draw[edge] (220) edge node[left]{$p\left(q_4, 0, 0\right) = \left(\frac{1}{\sqrt{2}}\right)^2$}  (230);
    	\draw[edge,cross] (220) edge node[right]{$...$}  (231);
    	\draw[edge,cross] (221) edge node[left]{$...$}  (230);
    	\draw[edge] (221) edge node[right]{$p\left(q_4, 1, 1\right) = \left(\frac{1}{\sqrt{2}}\right)^2$}  (231);
    	\end{tikzpicture}
    	\caption{ Left shows the edge weights, Right shows how to compute transition probabilities given edge weights for $\ket{\Psi}=\frac{1}{2}\left(\ket{0000}+ \ket{0001} + \ket{0010} + \ket{0011}\right)$ }
    	\label{fig:transprob}
    \end{figure}

    Method 1 outlines the \textit{Measure State} procedure. Running Method 1 once will produce a bit string of length $n$. To obtain a probability distribution that corresponds to our input state $\ket{\Psi}$ we run Method 1 many times and record the results in a frequency table. Note that line $10$ of the algorithm is equivalent to emitting the destination node label. The key feature of the ICM measurement process is expressed in equation $(2)$: \textit{next steps in the walk are conditionally independent of past steps provided that their current values are known.} Therefore, QuMvNs satisfy the \textit{Markov Property}. An example of how to compute transition probabilities given node location and edge weight is shown in FIG. \ref{fig:transprob}.
    
    After each qubit measurement, our knowledge of the quantum state $\ket{\Psi}$ changes, the so called ``collapse'' of the wave function. QuMvNs show that wave function collapse is equivalent to a state transition from the state at step $t$ to the state at step $t+1$: $\ket{\Psi^t} \rightarrow \ket{\Psi^{t+1}}$. The state transition upon measurement and the Markov Property of equation $(2)$ show that QuMvNs are Markov Chains.

    \subsection{The Wave Function as an Information Source}
    
    Shannon shows that a Markov Chain is a discrete information source if it is ergodic \cite{shannon1948mathematical}. Therefore, if we are able to make a single layer QuMvN ergodic, we have a discrete information source. 
    
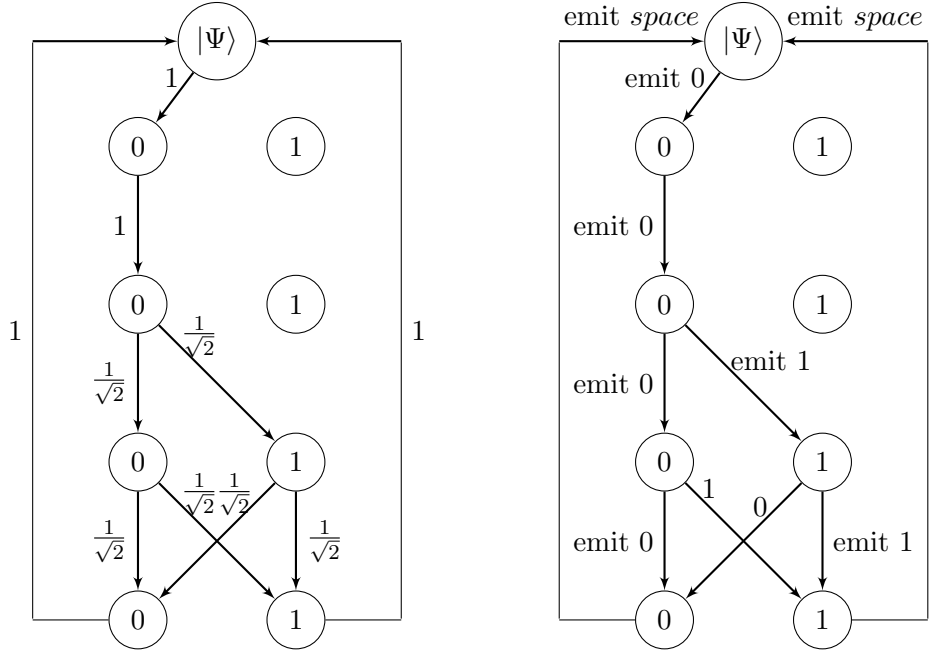
\begin{figure}[!h]
	\centering
	\begin{tikzpicture}[scale=0.7]
	\tikzset{every loop/.style={min distance=10mm,in=0,out=60,looseness=10}}
	\node[vertex] (r) at  (1.5,2) {$\ket{\Psi}$};
	
	\node[vertex] (00) at  (0,0) {$0$};
	\node[vertex] (01) at  (3,0) {$1$};	
	\node[vertex] (10) at  (0,-3) {$0$};
	\node[vertex] (11) at  (3,-3) {$1$};
	\node[vertex] (20) at  (0,-6) {$0$};
	\node[vertex] (21) at  (3, -6) {$1$};	
	\node[vertex] (30) at  (0,-9) {$0$};
	\node[vertex] (31) at  (3,-9) {$1$};
	\node[invisible] (il1) at (-2, -9) {};
	\node[invisible] (il2) at (-2, 2) {};
	\node[invisible] (ir1) at (5, -9) {};
	\node[invisible] (ir2) at (5, 2) {};
	
	\draw[edge, cross] (r) edge node[left]{$1$} (00);
	\draw[edge] (00) edge node[left]{$1$}  (10);
	\draw[edge,cross] (10) edge node[right]{$\frac{1}{\sqrt{2}}$}  (21);
	\draw[edge] (10) edge node[left]{$\frac{1}{\sqrt{2}}$}  (20);
	\draw[edge] (20) edge node[left]{$\frac{1}{\sqrt{2}}$}  (30);
	\draw[edge,cross] (20) edge node[right]{$\frac{1}{\sqrt{2}}$}  (31);
	\draw[edge,cross] (21) edge node[left]{$\frac{1}{\sqrt{2}}$}  (30);
	\draw[edge] (21) edge node[right]{$\frac{1}{\sqrt{2}}$}  (31);
	\draw (30) edge (il1);
	\draw (il1) edge node[left]{$1$} (il2);
	\draw[edge, left] (il2) edge (r);
	
	\draw (31) edge (ir1);
	\draw (ir1) edge node[right]{$1$} (ir2);
	\draw[edge, left] (ir2) edge (r);

	\node[vertex] (r2) at  (11.5,2) {$\ket{\Psi}$};
	
	\node[vertex] (200) at  (10,0) {$0$};
	\node[vertex] (201) at  (13,0) {$1$};	
	\node[vertex] (210) at  (10,-3) {$0$};
	\node[vertex] (211) at  (13,-3) {$1$};
	\node[vertex] (220) at  (10,-6) {$0$};
	\node[vertex] (221) at  (13, -6) {$1$};	
	\node[vertex] (230) at  (10,-9) {$0$};
	\node[vertex] (231) at  (13,-9) {$1$};
	\node[invisible] (2il1) at (8, -9) {};
	\node[invisible] (2il2) at (8, 2) {};
	\node[invisible] (2ir1) at (15, -9) {};
	\node[invisible] (2ir2) at (15, 2) {};
	
	\draw[edge, cross] (r2) edge node[left]{emit $0$} (200);
	\draw[edge] (200) edge node[left]{emit $0$}  (210);
	\draw[edge,cross] (210) edge node[right, style={pos=0.3}]{emit $1$}  (221);
	\draw[edge] (210) edge node[left]{emit $0$}  (220);
	\draw[edge] (220) edge node[left]{emit $0$}  (230);
	\draw[edge,cross] (220) edge node[right, style={pos=0.05}]{ $1$}  (231);
	\draw[edge,cross] (221) edge node[left, style={pos=0.2}]{$0$}  (230);
	\draw[edge] (221) edge node[right]{emit $1$}  (231);
	
	\draw (230) edge (2il1);
	\draw (2il1) edge node[left]{} (2il2);
	\draw[edge, left] (2il2) edge node[above]{emit $space$}(r2);
	
	\draw (231) edge (2ir1);
	\draw (2ir1) edge node[right]{} (2ir2);
	\draw[edge, left] (2ir2) edge node[above]{emit $space$} (r2);
	\end{tikzpicture}
	\caption{ Example: Left shows the edge weights, Right shows how to emit symbols and messages as an information source for $\ket{\Psi}=\frac{1}{2}\left(\ket{0000}+ \ket{0001} + \ket{0010} + \ket{0011}\right)$ }
	
	\label{fig:emit}
\end{figure}

    The current single layer QuMvN description is not ergodic as is. Rather, the source stops emitting information once qubit $n$ is reached during measurement, making it an absorbing Markov Chain. We can make single layer QuMvN ergodic by introducing a delimiter $d$  to our alphabet. For simplicity, let the delimiter be a $space$ character. Then, once we reach qubit $n$ we return to the root of our QuMvN with probability $1$. A transition to root emits the delimiter in our alphabet. An example is shown in FIG. \ref{fig:emit}. This can be viewed as running the measurement procedure many times. We keep performing measurements until the probability distribution converges, and it is guaranteed to converge if the Markov Chain representation of a single layer QuMvN is ergodic.
    
    Creating a path between the last qubit nodes and the root node $\ket{\Psi}$ gives root the following property: node $\ket{\Psi}$ is reachable from any other node in finite time. The above property along with the fact that starting from $\ket{\Psi}$ we can reach any possible state in $n$ steps implies that a single layer QuMvN satisfies the Doeblin Condition \cite{tweedie1975sufficient, stroock2013introduction} and has a stationary probability distribution as the number of measurements $\rightarrow \infty$. Therefore, a single layer QuMvN is ergodic.
    
    To summarize, a single layer QuMvN now has the following properties:
    
    \begin{enumerate}
    	\item an alphabet $A=\{0,1,space\}$ where $space$ can be any delimiter
    	\item it satisfies the Markov Condition thus is a Markov Chain
    	\item it is ergodic  
    	\item we emit a message by emitting $n$ characters from our alphabet along with a space according to the ICM with probabilities given in equation 2
    \end{enumerate}

	These four properties define a discrete information source as described by Shannon. The message length of the information source is $n$ (the number of qubits). Messages of the information source are possible measurement outcomes of the separable wave function it represents. Messages are generated by emitting characters from an alphabet as shown in FIG. \ref{fig:emit}. A separable wave function is a discrete information source and QuMvNs are an efficient representation of the information source.

    %\newpage
	\subsection{Single Qubit Gates}
	
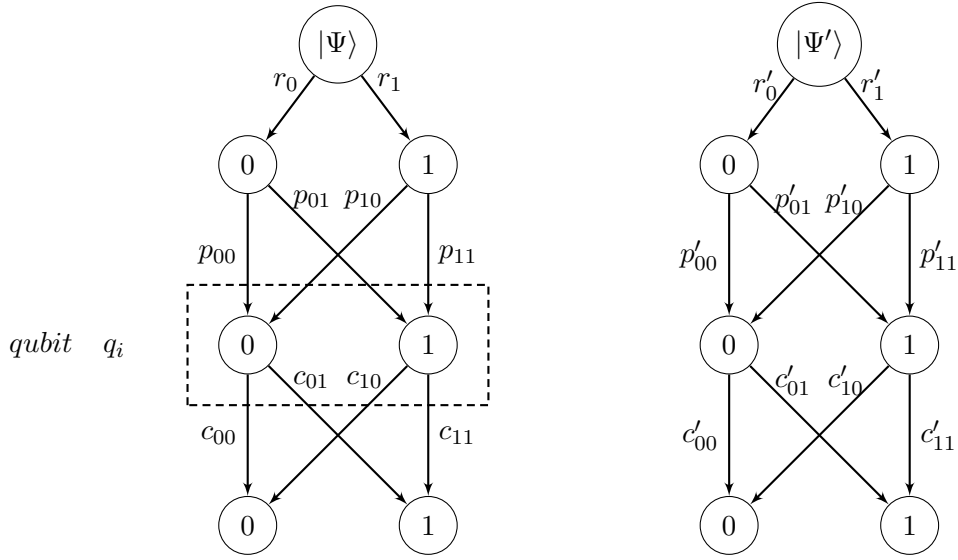
\begin{figure}[h]
	\centering
	\begin{tikzpicture}[scale=0.8]
	
	\node[vertex] (r) at  (1.5,2) {$\ket{\Psi}$};
	\node[vertex] (00) at  (0,0) {$0$};
	\node[vertex] (01) at  (3,0) {$1$};	
	
	\node[vertex] (10) at  (0,-3) {$0$};
	\node[vertex] (11) at  (3,-3) {$1$};	
	
	\node[vertex] (20) at  (0,-6) {$0$};
	\node[vertex] (21) at  (3,-6) {$1$};

	\draw[edge] (00) edge node[left]{$p_{00}$}  (10);
	\draw[edge,cross] (00) edge node[right]{$p_{01}$}  (11);
	
	\draw[edge, cross] (01) edge node[left]{$p_{10}$} (10);
	\draw[edge] (01) edge node[right]{$p_{11}$} (11);
	
	\draw[edge] (10) edge node[left]{$c_{00}$} (20);
	\draw[edge, cross] (10) edge node[right]{$c_{01}$} (21);
	
	\draw[edge, cross] (11) edge node[left]{$c_{10}$} (20);
	\draw[edge] (11) edge node[right]{$c_{11}$} (21);
	
	\draw[edge, cross] (r) edge node[left]{$r_0$} (00);
	\draw[edge, cross] (r) edge node[right]{$r_1$} (01);
	
	\draw[line, black, densely dashed] (-1,-4) rectangle (4,-2);
	\node (q2) at (-3,-3) {$qubit\phantom{a}\phantom{a}q_i$};
	
	\node[vertex] (r2) at  (9.5,2) {$\ket{\Psi'}$};
	\node[vertex] (00) at  (8,0) {$0$};
	\node[vertex] (01) at  (11,0) {$1$};	
	
	\node[vertex] (10) at  (8,-3) {$0$};
	\node[vertex] (11) at  (11,-3) {$1$};	
	
	\node[vertex] (20) at  (8,-6) {$0$};
	\node[vertex] (21) at  (11,-6) {$1$};

	\draw[edge] (00) edge node[left]{$p'_{00}$}  (10);
	\draw[edge,cross] (00) edge node[right]{$p'_{01}$}  (11);
	
	\draw[edge, cross] (01) edge node[left]{$p'_{10}$} (10);
	\draw[edge] (01) edge node[right]{$p'_{11}$} (11);
	
	\draw[edge] (10) edge node[left]{$c'_{00}$} (20);
	\draw[edge, cross] (10) edge node[right]{$c'_{01}$} (21);
	
	\draw[edge, cross] (r2) edge node[left]{$r'_0$} (00);
	\draw[edge, cross] (r2) edge node[right]{$r'_1$} (01);
	
	\draw[edge, cross] (11) edge node[left]{$c'_{10}$} (20);
	\draw[edge] (11) edge node[right]{$c'_{11}$} (21);	
	
	\end{tikzpicture}
	\caption{single qubit gate applied on qubit $q_i$}
	\label{fig:pc}
\end{figure}

	We now discuss how to begin with a initial state $\ket{0}^{\otimes n}$ and obtain a more complicated information source. QuMvNs allow for the efficient implementation of some single qubit gates by treating them as operations on the edge weights instead. Quantum Gates are implemented with specific rules on how to transform parent and child edge weights of sibling nodes in a QuMvN as outlined in FIG. \ref{fig:pc}. 
	
	The $X$, $H$, and $R(k)$ single qubit gates are described in detail, however, many variations and alterations to the following details may be used to create additional single qubit gates. With the equivalent of single qubit gates implemented on a QuMvN, we can start with the initial state $\ket{0}^{\otimes n}$ and obtain more complicated states by implementing quantum circuits.

Note that special care must be taken for the first qubit because the parent node for the first qubit is the \textit{root} node which is not another qubit node. The edges of the root node are labeled $r_0$ and $r_1$ for the edges directed towards the zero node and one node respectively. We show a separate description of the first qubit in our gates below for added clarity.

\subsubsection{X: NOT gate}

To apply the $X$ gate to qubit $q_i$ on a single layer QuMvN:

\begin{enumerate}
	\item go to qubit $q_i$
	
	\item swap child edge weights of siblings in qubit $q_i$:
	\begin{align*}
	c'_{00} &= c_{10}  \\
	c'_{01} &= c_{11} \\
	c'_{10} &= c_{00} \\
	c'_{11} &= c_{01}
	\end{align*}
	
	\item if $q_i=1$ (root node parent):
	\begin{align*}
	r'_0 = r_1 \\
	r'_1 = r_0
	\end{align*}	
	\item if $q_i \ne 1$ swap parent edge weights of siblings in qubit $q_i$: 
	\begin{align*}
	p'_{00} &= p_{01} \\
	p'_{10} &= p_{11} \\
	p'_{01} &= p_{00} \\
	p'_{11} &= p_{10}
	\end{align*}
\end{enumerate}

\subsubsection{H: Haddamard gate}

To apply the $H$ gate to qubit $q_i$ on a single layer QuMvN:
\begin{enumerate}
	\item go to qubit $q_i$
	
	\item if $q_i=1$ (root node parent):
	\begin{align*}
	r'_0 &= r_{0} \left( \frac{1}{\sqrt{2}} \right) + r_{1}  \left( \frac{1}{\sqrt{2}} \right) \\ 
	r'_1 &= r_{0} \left( \frac{1}{\sqrt{2}} \right) + r_{1}  \left( \frac{-1}{\sqrt{2}} \right) 
	\end{align*}	
	
	\item if $q_i \ne 1$ interfere parent edges in qubit layer $q$:
	\begin{align*}
	p'_{00} &= p_{00} \left( \frac{1}{\sqrt{2}} \right) + p_{01}  \left( \frac{1}{\sqrt{2}} \right) \\
	p'_{10} &=  p_{10} \left( \frac{1}{\sqrt{2}} \right) + p_{11}  \left( \frac{1}{\sqrt{2}} \right)\\
	p'_{01} &=  p_{00} \left( \frac{1}{\sqrt{2}} \right) + p_{01}  \left( \frac{-1}{\sqrt{2}} \right) \\
	p'_{11} &=  p_{10} \left( \frac{1}{\sqrt{2}} \right) + p_{11}  \left( \frac{-1}{\sqrt{2}} \right)		
	\end{align*}
	
	\item copy nonzero child edges: 
	\begin{itemize}
		\item if $c_{00} = 0$ and $c_{10} \ne 0$ or $c_{00}  \ne 0$ and $c_{10} = 0$ then set $c_{00} = c_{10}$
		\item if $c_{01} = 0$ and $c_{11} \ne 0$ or $c_{01}  \ne 0$ and $c_{11} = 0$ then set $c_{01} = c_{11}$
	\end{itemize}
\end{enumerate}

\subsubsection{R(k): phase gates}

$R(k)$ gates are the family of phase gates. For each $k \ge 0$ we have a distinct gate $R(k)$. The following method applies to the family of phase gates:

$$
R(k) = 
\begin{bmatrix}
1 & 0 \\
0 & \phi(k)
\end{bmatrix}
$$

where $\phi(k)$ is the \textit{phase} defined as $$\phi(k) = e^{\frac{2 \pi i}{2^{k}}}$$

To apply the $R(k)$ gate to qubit number $q_i$ on a single layer QuMvN: 
\begin{enumerate}
	\item go to qubit $q_i$
	
	\item if $q_i=1$ (root node parent):
	\begin{align*}
	r'_1 = r_1 * \phi(k)
	\end{align*}	
	
	\item if $q_i \ne 1$ apply the phase to parent edges of the 1 node:
	\begin{align*}
	p'_{01} &= p_{01} * \phi(k) \\
	p'_{11} &= p_{11} * \phi(k)
	\end{align*}
\end{enumerate}
	
	\section{Entanglement: Multilayer QuMvN}
	
	To extend our description to entangled states we use multiple layers in a multiplex network and create an $L$ layer QuMvN where $L$ is the number of layers. Suppose $\ket{Q}$ is an entangled state, then by definition we cannot write $\ket{Q}$ as a tensor product of states. However, if we decompose our wave function $\ket{Q}$ as a sum of separable components then our wave function resembles what Shannon referred to as a ``mixed'' information source \cite{shannon1948mathematical} which is a linear combination of discrete information sources. Let $\Psi_1, \Psi_2, ... , \Psi_L$ be separable components of the wave function $\ket{Q}$, then we write $\ket{Q} = \omega_1\Psi_1 + \omega_2\Psi_2 + ... + \omega_L\Psi_L$ where $\omega_i \in \mathbb{C}$ and the probability of measuring a state from layer $\Psi_i$ is $\left|\omega_i\right|^2$. Note that the decomposition of $\ket{Q}$ is not necessarily unique and always admits a trivial decomposition where each separable component is just a basis vector in the corresponding Hilbert Space.

	\subsection{Multiplex Structure}
	A complete guide to multilayer networks is given by Boccaletti \cite{boccaletti2014structure}, however, QuMvNs are specific types of multilayer networks called a \textit{multiplex network}. In a multiplex network each layer has the same nodes. Each layer of a QuMvN consists of a separable component of a wave function $\ket{\Psi_i}$ represented by an ergodic Markov Chain (as outlined in Section 2). Take for example the Bell State $\ket{Q_1} = \frac{1}{\sqrt{2}}\ket{00} + \frac{1}{\sqrt{2}}\ket{11}$. We cannot represent $\ket{Q_1}$ as a tensor product of states therefore no single layer QuMvN representation exists. However, there is a two layer QuMvN representation $\ket{Q_1} = \frac{1}{\sqrt{2}}\Psi_1 + \frac{1}{\sqrt{2}}\Psi_2$ where $\Psi_1=\ket{00}$, $\Psi_2=\ket{11}$ as depicted in FIG. \ref{fig:bell}.

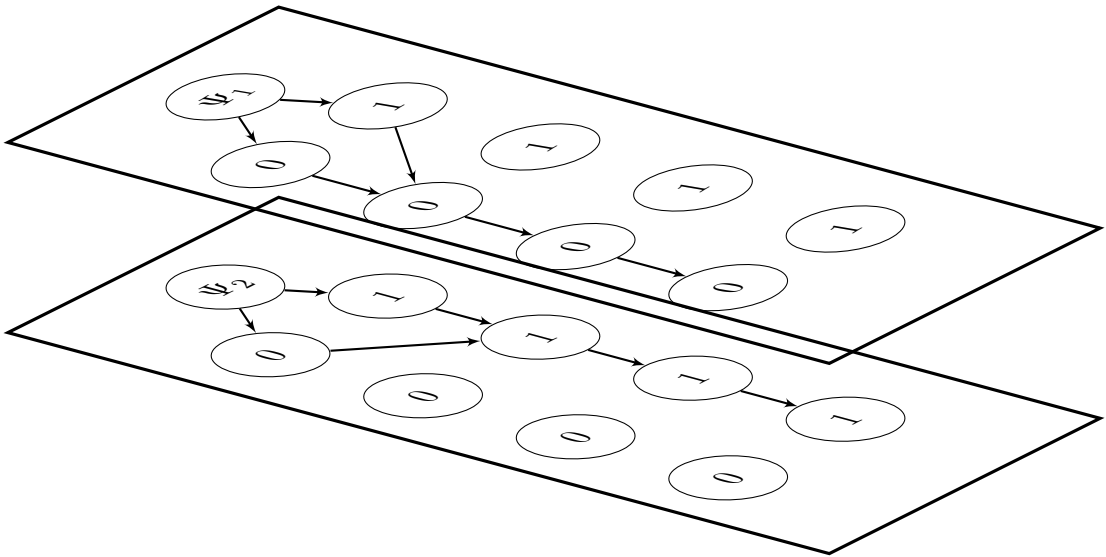
\begin{figure}[h]
	\centering
	\begin{tikzpicture}[scale=1.2,every node/.style={shape=circle,draw,minimum size=2.5em},on grid]
	
	% slanting: production of a set of n 'laminae' to be piled up.
	% N=number of grids.
	\begin{scope}[
	yshift=-100,every node/.append style={
		yslant=0.6,xslant=-1.3},yslant=0.5,xslant=-1.3
	]
	% opacity to prevent graphical interference
	\fill[white,fill opacity=0.9] (1,0) rectangle (4,7);
	\draw[black,very thick] (1,0) rectangle (4,7);%marking borders      

	\node (L1) at (2.5, 6.3) {$\Psi_1$} ;
	\node (L100) at (1.7, 5.3) {$0$} ;
	\node (L101) at (3, 5.3) {$1$} ;
	\node (L110) at (1.7, 4.0) {$0$} ;
	\node (L111) at (3, 4.0) {$1$} ;
	\node (L120) at (1.7, 2.7) {$0$} ;
	\node (L121) at (3, 2.7) {$1$} ;
	\node (L130) at (1.7, 1.4) {$0$} ;
	\node (L131) at (3, 1.4) {$1$} ;
	
	\draw[edge] (L1) edge  (L100);
	\draw[edge] (L1) edge (L101);
	\draw[edge] (L100) edge (L110);
	\draw[edge] (L101) edge (L110);
	\draw[edge] (L110) edge (L120);
	\draw[edge] (L120) edge (L130);

	\pgfkeys{/pgf/number format/.cd, fixed, zerofill, precision =1}

	\end{scope}
	
	\begin{scope}[
	yshift=-160,every node/.append style={
		yslant=0.5,xslant=-1.3},yslant=0.5,xslant=-1.3
	]
	%marking border
	\draw[black,very thick] (1,0) rectangle (4,7);

	%draw bottom parabola
	
	\node (L2) at (2.5, 6.3) {$\Psi_2$} ;
\node (L200) at (1.7, 5.3) {$0$} ;
\node (L201) at (3, 5.3) {$1$} ;
\node (L210) at (1.7, 4.0) {$0$} ;
\node (L211) at (3, 4.0) {$1$} ;
\node (L220) at (1.7, 2.7) {$0$} ;
\node (L221) at (3, 2.7) {$1$} ;
\node (L230) at (1.7, 1.4) {$0$} ; 
\node (L231) at (3, 1.4) {$1$} ;

\draw[edge] (L2) edge  (L200);
\draw[edge] (L2) edge (L201);
\draw[edge] (L200) edge (L211);
\draw[edge] (L201) edge (L211);
\draw[edge] (L211) edge (L221);
\draw[edge] (L221) edge (L231);

	\end{scope} %end of drawing grids
	
	% s
	%\draw[-latex,thick](-4,-.2)node[left,scale=1.3]{$\phi_\Omega(s)$}
	%to[out=0,in=90] (sphi);
	
	\end{tikzpicture}
	\caption{$\ket{Q_2} = \frac{1}{\sqrt{2}}\Psi_1 + \frac{1}{\sqrt{2}}\Psi_2 $}
	\label{fig:q2}
\end{figure}

The Bell State is a simple QuMvN representation. A more complicated QuMvN is the four qubit state $\ket{Q_2} = \frac{1}{2} \left(\ket{0000} + \ket{0111} + \ket{1000} + \ket{1111}\right)$. The trivial representation requires four layers with each layer corresponding to a possible vector in the $\ket{Q_2}$ superposition. However, a better representation exists. Let $\Psi_1=\left(\frac{1}{\sqrt{2}} \ket{0} + \frac{1}{\sqrt{2}} \ket{1}\right) \otimes \ket{000}$ and $\Psi_2=\left(\frac{1}{\sqrt{2}} \ket{0} + \frac{1}{\sqrt{2}} \ket{1}\right) \otimes \ket{111}$ then $\ket{Q_2} = \frac{1}{\sqrt{2}}\Psi_1 + \frac{1}{\sqrt{2}}\Psi_2 $ which requires fewer resources to simulate classically FIG. \ref{fig:q2}. Single qubit gates are applied to a qubit in each layer. Therefore, the complexity of applying single qubit gates is now $\mathcal{O}\left(nL\right)$ for $n$ qubits and $L$ layers. Given a random wave function and determining the \textit{minimum} number of layers required to represented it is a computationally difficult problem \cite{de2015structural}. However, some quantum circuits have less complicated layer structures. We found that Shor's Algorithm is such a circuit as explained in Section $3.3$.

	\subsection{Creating layers with control gates}
	Layers in a QuMvN are created by control gates. We always initialize our state in the initial state $\ket{G} = \ket{0}^{\otimes n}$ and apply quantum logic gates as outlined in Section 2.4. We only need one layer to represent the initial state, we define this layer as $\Psi_0$ with coefficient $\omega_0 = 1$. Applying single qubit gates to $\ket{G} = \omega_0\Psi_0 = \ket{0}^{\otimes n}$ does not create any additional layers, but once a control gate is applied new layers may be created. 
	
	Let $U$ be a single qubit gate defined as an edge transformation as in Section 2.4. Let control-$U$ be a two qubit operation where $q_i$ is the control qubit and $q_j$ the target qubit. At a high level, a control-$U$ operation creates a new layer for edges that pass through the 1 node of $q_i$, then applies $U$ to $q_j$ of that layer. Details are given in Method 2 listing.
	
	As an example, we generate the Bell State in FIG. \ref{fig:bell} by using a $CNOT$ gate. We begin with the initial state $\ket{00}$ and apply a $H$ to the first qubit which gives $\ket{Q_1} = \frac{1}{\sqrt{2}}\ket{00} + \frac{1}{\sqrt{2}}\ket{10}$. $\ket{Q_1}$ is a separable state thus requires only one layer depicted in FIG. \ref{fig:bell1}. Then we apply an $X$ gate on qubit $2$ controlled by qubit $1$. First, split the single layer into two layers $\Psi_1, \Psi_2$ with coefficients $\omega_1 = \omega_2 = \frac{1}{\sqrt{2}}$ where the control qubit is $0$ in $\Psi_1$ and $1$ in $\Psi_2$ as shown in FIG. \ref{fig:bell2}. Lastly, apply the $X$ gate in the layer where the control qubit is $1$ and obtain the Bell state $\ket{Q_1} = \frac{1}{\sqrt{2}}\ket{00} + \frac{1}{\sqrt{2}}\ket{11}$ in FIG. \ref{fig:bell}

	\begin{figure}[h]
		\centering
		\begin{tikzpicture}[scale=1.1,every node/.style={shape=circle,draw,minimum size=2.5em},on grid]
		
		% slanting: production of a set of n 'laminae' to be piled up.
		% N=number of grids.
		\begin{scope}[
		yshift=-100,every node/.append style={
			yslant=0.6,xslant=-1.3},yslant=0.5,xslant=-1.3
		]
		% opacity to prevent graphical interference
		\fill[white,fill opacity=0.9] (1,0) rectangle (4,4);
		\draw[black,very thick] (1,0) rectangle (4,4);%marking borders      

		\node (L1) at (2.5, 3.3) {$\Psi_1$} ;
		\node (L100) at (1.7, 2.3) {$0$} ;
		\node (L101) at (3, 2.3) {$1$} ;
		\node (L110) at (1.7, 1.0) {$0$} ;
		\node (L111) at (3, 1.0) {$1$} ;
		
		\draw[edge] (L1) edge  (L100);
		\draw[edge] (L100) edge (L110);
		\draw[edge] (L1) edge (L101);
		\draw[edge] (L101) edge (L110);

		\pgfkeys{/pgf/number format/.cd, fixed, zerofill, precision =1}

		\end{scope}
		
		\end{tikzpicture}
		\caption{$\ket{Q_1} = \frac{1}{\sqrt{2}}\ket{00} + \frac{1}{\sqrt{2}}\ket{10}$}
		\label{fig:bell1}
	\end{figure}
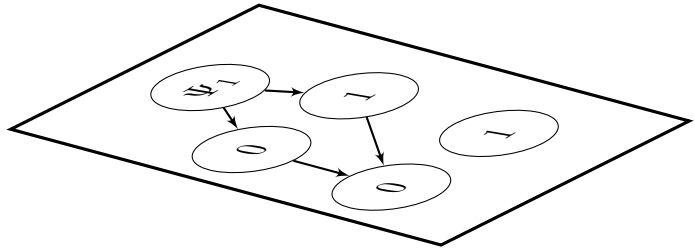

\begin{figure}[h]
	\centering
	\begin{tikzpicture}[scale=1.1,every node/.style={shape=circle,draw,minimum size=2.5em},on grid]
	
	% slanting: production of a set of n 'laminae' to be piled up.
	% N=number of grids.
	\begin{scope}[
	yshift=-100,every node/.append style={
		yslant=0.6,xslant=-1.3},yslant=0.5,xslant=-1.3
	]
	% opacity to prevent graphical interference
	\fill[white,fill opacity=0.9] (1,0) rectangle (4,4);
	\draw[black,very thick] (1,0) rectangle (4,4);%marking borders      

	\node (L1) at (2.5, 3.3) {$\Psi_1$} ;
	\node (L100) at (1.7, 2.3) {$0$} ;
	\node (L101) at (3, 2.3) {$1$} ;
	\node (L110) at (1.7, 1.0) {$0$} ;
	\node (L111) at (3, 1.0) {$1$} ;
	
	\draw[edge] (L1) edge  (L100);
	\draw[edge] (L100) edge (L110);

	\pgfkeys{/pgf/number format/.cd, fixed, zerofill, precision =1}

	\end{scope}
	
	\begin{scope}[
	yshift=-160,every node/.append style={
		yslant=0.5,xslant=-1.3},yslant=0.5,xslant=-1.3
	]
	%marking border
	\draw[black,very thick] (1,0) rectangle (4,4);

	%draw bottom parabola
	
	\node (L2) at (2.5, 3.3) {$\Psi_2$} ; 
	\node (L200) at (1.7,2.3) {$0$} ;
	\node (L201) at (3,2.3) {$1$} ;
	\node (L210) at (1.7, 1.0) {$0$};
	\node (L211) at (3, 1.0) {$1$} ;
	
	\draw[edge] (L2) edge  (L201);
	\draw[edge] (L201) edge (L210);

	\end{scope} %end of drawing grids
	
	% s
	%\draw[-latex,thick](-4,-.2)node[left,scale=1.3]{$\phi_\Omega(s)$}
	%to[out=0,in=90] (sphi);
	
	\end{tikzpicture}
	\caption{$\ket{Q_1}=\omega_1\ket{00} + \omega_2\ket{10}$}
	\label{fig:bell2}
\end{figure}
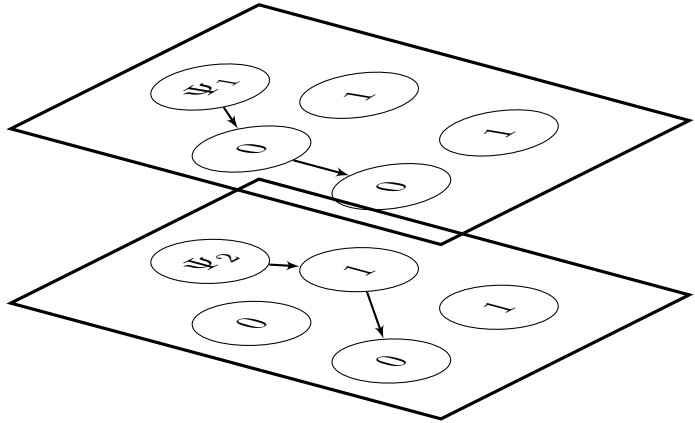

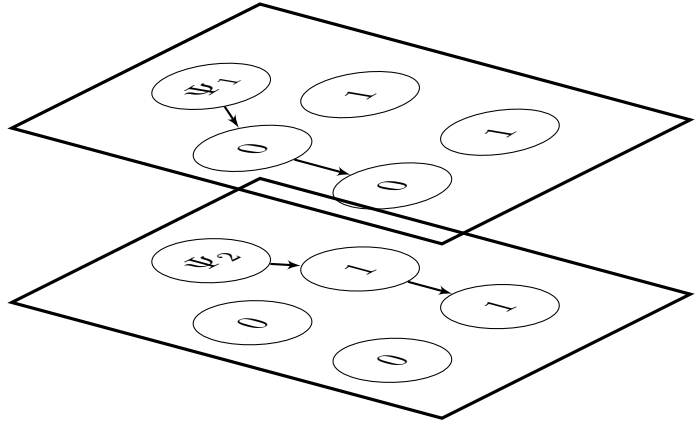
\begin{figure}[h]
	\centering
	\begin{tikzpicture}[scale=1.1,every node/.style={shape=circle,draw,minimum size=2.5em},on grid]
	
	% slanting: production of a set of n 'laminae' to be piled up.
	% N=number of grids.
	\begin{scope}[
	yshift=-100,every node/.append style={
		yslant=0.6,xslant=-1.3},yslant=0.5,xslant=-1.3
	]
	% opacity to prevent graphical interference
	\fill[white,fill opacity=0.9] (1,0) rectangle (4,4);
	\draw[black,very thick] (1,0) rectangle (4,4);%marking borders      

	\node (L1) at (2.5, 3.3) {$\Psi_1$} ;
	\node (L100) at (1.7, 2.3) {$0$} ;
	\node (L101) at (3, 2.3) {$1$} ;
	\node (L110) at (1.7, 1.0) {$0$} ;
	\node (L111) at (3, 1.0) {$1$} ;
	
	\draw[edge] (L1) edge  (L100);
	\draw[edge] (L100) edge (L110);

	\pgfkeys{/pgf/number format/.cd, fixed, zerofill, precision =1}

	\end{scope}
	
	\begin{scope}[
	yshift=-160,every node/.append style={
		yslant=0.5,xslant=-1.3},yslant=0.5,xslant=-1.3
	]
	%marking border
	\draw[black,very thick] (1,0) rectangle (4,4);

	%draw bottom parabola
	
	\node (L2) at (2.5, 3.3) {$\Psi_2$} ; 
	\node (L200) at (1.7,2.3) {$0$} ;
	\node (L201) at (3,2.3) {$1$} ;
	\node (L210) at (1.7, 1.0) {$0$};
	\node (L211) at (3, 1.0) {$1$} ;
	
	\draw[edge] (L2) edge  (L201);
	\draw[edge] (L201) edge (L211);

	\end{scope} %end of drawing grids
	
	% s
	%\draw[-latex,thick](-4,-.2)node[left,scale=1.3]{$\phi_\Omega(s)$}
	%to[out=0,in=90] (sphi);
	
	\end{tikzpicture}
	\caption{Two layer QuMvN for Bell State}
	\label{fig:bell}
\end{figure}

		\begin{algorithm}
			\floatname{algorithm}{Method}
		\caption{control-$U$ operation (also see Figure \ref{fig:pc})}
		\begin{algorithmic}[1]
			\Procedure{control($U$)}{}
			\For{$\Psi_i$ in all Layers}

			\If{$P\left(q_i = \ket{1}\right)$ != 0 and $P\left(q_i = \ket{0}\right)$ != 0}
				\State{create new layer $\Psi_{i'}$ $\gets$$\Psi_i$}
				\State{// let the degree of node $0$ and $1$ be $d(0),d(1)$ respectively}				

					\State{$\omega_{i'} \gets \omega_{i} \times \frac{p_{01}+p_{11}}{d(1)}$}
					
					\State{$\omega_{i} \gets \omega_{i} \times \frac{p_{00}+p_{10}}{d(0)}$}
				\State{// In the $\Psi_{i'}$ layer we pass through node $1$ of $q_i$ with certainty}			
				\State{$p_{01} \gets p_{11} \gets 1$ in layer $\Psi_{i'}$}
				\State{$p_{00} \gets p_{10} \gets 0$ in layer $\Psi_{i'}$}
				
				\State{// In $\Psi_i$ layer we pass through node $0$ of $q_i$ with certainty}
				\State{$p_{01} \gets p_{11} \gets 0$ in layer $\Psi_{i}$}
				\State{$p_{00} \gets p_{10} \gets 1$ in layer $\Psi_{i}$}
				\State{apply $U$ to $q_j$ in layer $\Psi_{i'}$}
			\Else
				\If{$P\left(q_i = \ket{1}\right)$ != 0}
					\State{apply $U$ to $q_j$ in layer $\Psi_{i}$}	
				\EndIf	
			\EndIf
			\EndFor
			\EndProcedure
		\end{algorithmic}
	\end{algorithm}

	\clearpage
	\subsection{Quantum Worlds, Network Layers and Complexity}
	
	The QuMvN formalism makes a connection between a wave function and a ``mixed'' information source consisting of many discrete information sources as described by Shannon. Each discrete information source is represented as a layer in the QuMvN formulation. However, what is the physical interpretation of the network layers? We find the most convenient way to think about the layers is in the context of the Many Worlds Interpretation \cite{everett1957relative} (MWI). In the MWI a network layer would be equivalent to a world. When applying a control-NOT gate, in one world the control qubit is $\ket{0}$ and in another world the qubit is $\ket{1}$. The NOT is applied in the world where the control is $\ket{1}$. Thinking about network layers of a QuMvN as a world in the MWI may give us a way to analyze the complexity of a wave function and the quantum circuit that generates the wave function. For the remainder of this paper, layers of QuMvNs are equivalent to worlds in the MWI.
	
	By complexity of a wave function $\ket{Q}$, we mean the difficulty to classically simulate the quantum circuit that takes the initial state $\ket{0}^{\otimes n}$ to the quantum state $\ket{Q}$. The main bottleneck in the classical simulation of quantum circuits in the QuMvN formalism is the number of worlds required to represent the quantum state $\ket{Q}$. For example, the quantum state in Figure \ref{fig:f4} has an exponential number of possible measurement outcomes. However, because there is no entanglement, only one world is needed to represent the wave function and is easily simulated classically using QuMvNs. Therefore, the quantity needed to measure complexity in the QuMvN formalism is the number of worlds needed to represent the wave function. 
	
	Each time a control-$U$ gate is applied to a control qubit in a superposition, we create new worlds. As the quantum computation progresses in a circuit do the layers remain decoherent? Are they separate worlds? The answer is generally no. Layers can interact and it is this interaction that determines which circuits can be efficiently simulated on a classical computer. By interaction we mean that layers may have the same state but different probability amplitudes for that state, thus they interfere. For example, if we have two layers such that layer one has the state $\phi_1\ket{010}$ and layer two has the state $-\phi_1\ket{010}$ then the states in those layers destructively interfere. The measurement method in a multilayer representation has to be modified depending on the structure of the layers. 
	
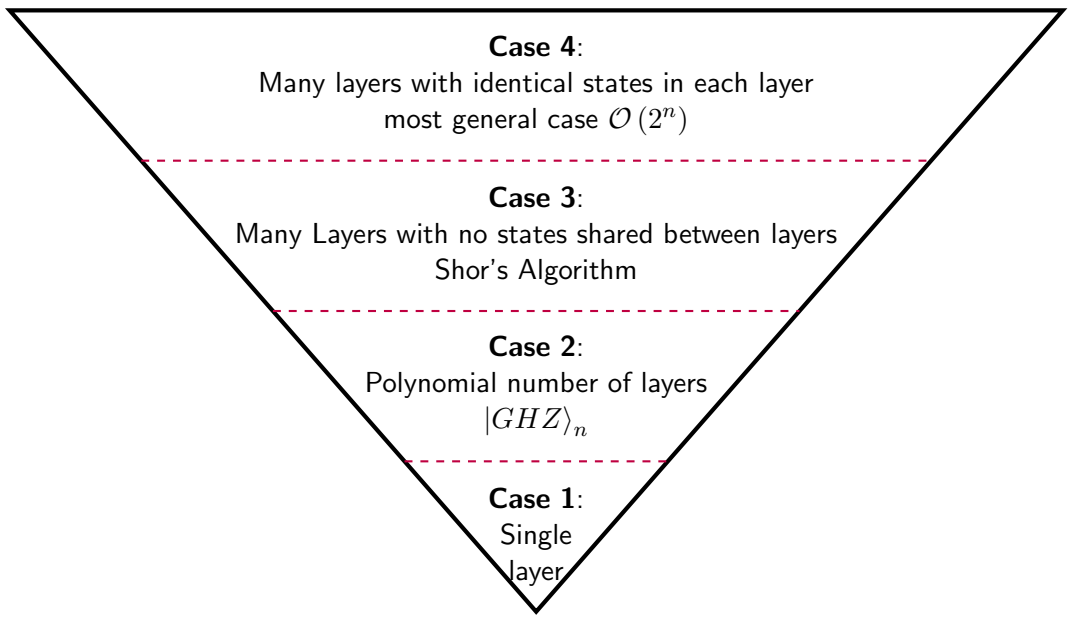
\begin{figure}[]
	\centering
	\caption{Layer Interaction Hierarchy}
\begin{tikzpicture}[font=\sffamily]
\draw[ultra thick] (0,0) -- (14,0) -- (7,-8) -- cycle;
\foreach \X in {1,...,3} {
	\draw[dashed, purple, thick] ({7*\X/4},{-8*\X/4}) -- ({14-7*\X/4},{-8*\X/4});
}

\node[text width=\dimexpr(6em)*5\relax,align=center] at (7, {-7}) {\textbf{Case 1}: \\ Single\\ layer};
\node[text width=\dimexpr(6em)*5\relax,align=center] at (7,-5) {\textbf{Case 2}: \\Polynomial number of layers \\ $\ket{GHZ}_n$};
\node[text width=\dimexpr(6em)*5\relax,align=center] at (7,-3) { \textbf{Case 3}: \\Many Layers with no states shared between layers \\ Shor's Algorithm};
\node[text width=\dimexpr(6em)*5\relax,align=center] at (7,-1) { \textbf{Case 4}: \\Many layers with identical states in each layer \\ most general case $\mathcal{O}\left(2^n\right)$};
\end{tikzpicture}
\label{fig:pyramid}
\end{figure}

	We implemented the QuMvN circuit simulation paradigm by developing a programmable virtual machine that executes quantum instructions backed by QuMvNs \cite{sakkaris2018}. Using our implementation we found patterns in the structure of the Layers by simulating random IQP circuits \cite{bremner2010classical} and Shor's Algorithm \cite{shor1999polynomial}. One surprising result that we observed is that Shor's Algorithm has a particular structure to the layers created which allowed us to simulate Shor's Algorithm on a classical computer at an impressive scale: $70$-qubits. We explore four different cases of layer structure and see that they form a hierarchy  depicted in FIG.\ref{fig:pyramid}:
	
	\begin{itemize}
		\item \textbf{Case 1 Single Layer:} QuMvNs with only one layer were explained in detail above with measurement outlined in Method 1. Circuits consisting of only single qubit gates outlined in Section 2.4 and semi-classical control gates are all single layer QuMvNs and can be efficiently simulated on a classical computer as expected. This may seem as an artificial problem, however, the terminal Quantum Fourier Transform is semi-classical ~\cite{griffiths1996semiclassical} and can be efficiently simulated on a classical computer with a single layer QuMvN ~\cite{sakkaris2016qudot}. In this case, no modification of the measurement method is needed.
		
		\item \textbf{Case 2 Polynomial Unique Layers:} one step after single layer QuMvNs is where the number of layers remains polynomial in the number of qubits \textit{and} each layer has a unique set of states. Detecting if layers represent unique states can be done in a variety of methods such as examining their adjacency matrix. Restricting each layer to a unique set of states allows us to modify our measurement Method 1 for multiple layers. Suppose we have $L$ layers with coefficients given by $$\ket{Q} = \omega_1\Psi_1 + \omega_2\Psi_2 + ... + \omega_L\Psi_L$$ To measure a state we first measure a layer with probability $P\left(\Psi_L\right)=\left|\omega_L\right|^2$ and then measure a state according to Method 1. The measurement procedure is now $\mathcal{O}\left(nL\right)$ for $n$ qubits and $L$ layers. If $L$ is polynomial in $n$ then Case 2 is polynomial in the number of qubits and can be efficiently simulated on a classical computer. An example of Case 2 is the circuit that generates the entangled $\ket{GHZ}_n$ states which only require two layers and the circuit that generates the state $\ket{Q_2}$ depicted in FIG. 11. 
		
		\item \textbf{Case 3 Exponential Unique Layers:} the third step of the layer interaction hierarchy is the same as Case 2, but the number of layers grows exponentially in the number of qubits. In this case we can use the same measurement procedure as in Case 2 but we cannot efficiently simulate on a classical computer. However, the QuMvN approach still has advantages over traditional matrix simulations that require $2^n \times 2^n$ matrix operations on vectors of size $2^n$. To our surprise we found that Shor's Algorithm ~\cite{shor1999polynomial} falls in Case 3. We showcase the benefits of the QuMvN simulation approach over the matrix mechanics approach by simulating Shor's Algorithm to $70$ qubits on a cloud server. To our knowledge this is the largest simulation of Shor's Algorithm available. Details are given in Section 4.
		
		\item \textbf{Case 4 Multiple Non-Unique Layers:} having multiple layers with the same states is the most general case in random IQP circuits. In this case we cannot use the measurement procedures discussed in Cases 1-3. To calculate the probability of a state we calculate the probability amplitude of a state in each layer, sum the probability amplitudes and then absolute square the sum. Since enumerating the paths in each layer is exponential in the number of qubits, the general case is exponentially difficult to simulate on a classical computer (even for polynomial layers).
	\end{itemize}

	 The layer interaction hierarchy of QuMvNs gives us a new tool to analyze quantum computing algorithms. If a quantum algorithm is analyzed and shown to be in Case 3, then we can simulate on classical computers to a respectable number of qubits and test our idea before running on expensive quantum hardware. We have successfully done such a simulation for Shor's Algorithm up to $70$-qubits as discussed in Section 4. If a quantum algorithm is shown to be in Case 1-2 then we can get the same speedup as quantum hardware by using the QuMvN formalism. This presents a distinct advantage over the matrix simulation of quantum circuits and gives us hints on the ``elusive source of quantum effectiveness'' \cite{vedral2010elusive}.
	
	There are three main hypothesis given to the source of quantum speedups as explained by Rieffel \cite{rieffel2011quantum}. The QuMvN point of view helps us better understand them:
	
	\begin{enumerate}
		\item Exponential size of the state space: a quantum system with $n$ qubits has $2^n$ possible states. However, this is not always a problem to simulate classically. For example, the $n$ qubit superposition in FIG. \ref{fig:f4} can easily be simulated using QuMvNs in $\mathcal{O}(n)$ using one layer because it is completely separable. Thus, the exponential size of the state space cannot be the source of quantum speedups on it's own.
		
		\item Parallelism: another popular explanation of quantum speedups is that quantum computers perform an exponential amount of computation in \textit{parallel}. The physical mechanism that accomplishes this amazing feat is not explained, it is obtained from the abstract mathematics of unitary operators. However, QuMvNs shed light on this mechanism as well. Consider FIG. \ref{fig:f4}, suppose we change the weight of a single edge in the QuMvN. We just affected any measurement that passes through that edge with a single operation. If that edge is one of the root edges, then we just affected an exponential number of possible outcomes using one simple classical operation. Therefore, the massive parallelism cannot be the only source of quantum speedups.
		
		\item Entanglement: it has been shown that any exponential speedup of quantum algorithms over classical makes use of entanglement that increases with the size of our input \cite{jozsa2003role}. QuMvNs show that entanglement adds layers to a QuMvN therefore increasing our simulation costs. At the same time, the QuMvN formalism suggests that entanglement creates layers with unique states. So it seems, entanglement can make classical simulation more difficult by adding layers but if entanglement increases to the extent where number of layers added is polynomial in qubit size, then we can efficiently simulate classically. For example, the circuit that generates the $\ket{GHZ_n} = \frac{1}{\sqrt{2}}\ket{000...0} + \frac{1}{\sqrt{2}}\ket{111...1}$ state is easily simulated with QuMvNs. We only need two layers, in one layer we eventually have the $\ket{000...0}$ state an in another layer we eventually have the $\ket{111...1}$ state. 
		
	\end{enumerate}

	\section{Simulation Results of Shor's Algorithm}
	
	To demonstrate the effectiveness of QuMvNs as a classical simulation technology, we developed a quantum bytecode virtual machine that is backed by QuMvNs. The KratosVM is a C++17 implementation of the QuDot Virtual Machine instruction set capable of performing Shor's Algorithm \cite{shor1999polynomial}. Details of the QuDot Virtual Machine instruction set and architecture can be found in \cite{sakkaris2018}. We used the Intel Thread Building Blocks library for parallelization and the Intel MKL library for random number generation. The implementation of Shor's algorithms is based on quantum arithmetic  components outlined in \cite{vedral1996quantum,van2005fast,beckman1996efficient}. We performed six simulations of Shor's Algorithm using $35$, $40$, $50$, $55$, $60$ and $70$ qubits factoring the semi-prime numbers $77$, $145$, $731$, $1273$, $2291$ and $10057$ respectively. All data generated by the KratosVM and classical post-processing script used to analyze the data is publicly available on GitHub \cite{sakkaris2019Kratos01}.
	
\subsection{Overview Of Shor's Algorithm}
Shor's Algorithm \cite{shor1999polynomial} is a famous result in quantum computation. Given an odd, semi-prime number $N=pq$, Shor's Algorithm can find the factors $p,q$ with high probability in polynomial time on a quantum computer. Besides the application to cryptography, Shor's Algorithm uses a mix of classical computing and quantum computing which makes it a great example of how we can potentially offload subroutines to quantum hardware in a heterogeneous computing systems. Also, Shor's Algorithm has the appealing property that the solutions are easily checked for correctness: we just multiply the two factors returned $p \times q$ and check if they equal our input $N$. We follow the description of Shor's algorithm given in \cite{rieffel2011quantum}. We first choose a random number $a \in [1, N-1]$. If $gcd(a, N) \neq 1$ then we have found a factor and are done (this is exponentially unlikely). Otherwise, we pass $a$ and $N$ into the quantum period finding algorithm which returns the period, $r$, of the function $f\left(x\right) = ax$ mod $N$. Lastly, $r$ can be used to find the factors $p,q$ with high probability using classical analysis as explained in detail by \cite{rieffel2011quantum, shor1999polynomial}.

\begin{figure}[h!]
	\includegraphics[width=\linewidth]{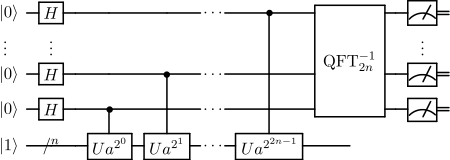}
	\caption{Circuit for Shor's Algorithm}
	\label{fig:shor1}
\end{figure}

The circuit of Shor's Algorithm uses two registers: an upper register with $k$ qubits initialized to $\ket{0}^{\otimes k}$ and a lower register with $n$ qubits initialized to $\ket{1}$. The number $n$ is determined by the number of bits required to store $N$ which is $\lceil log_2\left(N\right) \rceil$ and $k$ satisfies $N^2 \le 2^k < 2N^2$ (typically $k = 2n$ or $k=2n-1$). We show a circuit of Shor's Algorithm in FIG.~\ref{fig:shor1}, for a deeper explanation on the construction of FIG.~\ref{fig:shor1} we refer to the excellent explanations in \cite{rieffel2011quantum, nielsen2002quantum} 

In addition to the upper and lower register qubits, Shor's Algorithm requires additional qubits for arithmetic, such as, modular addition and modular multiplication \cite{beckman1996efficient}. For a lower register of $n$ qubits, the KratosVM requires an additional $2n+1$ qubits for arithmetic operations. These arithmetic qubits are managed by the virtual machine. The qubit requirements for our six simulation our outlined in Table \ref{tab:ShorQubits}.

\begin{table}[h!]
	\centering
	\caption{Qubit Counts For Shor's Algorithm}
	\label{tab:ShorQubits}
	\begin{tabular}{r|r|r|r|r|r}
		\textbf{$N$} & \textbf{$n$} & \textbf{$k$} & \textbf{$2n+1$} & \textbf{total}\\		
		\hline
		77 & 7 & 13 & 15 & 35 qubits \\
		145 & 8 & 15 & 17 & 40 qubits \\
		731 & 10 & 19 & 21 & 50 qubits \\
		1273 & 11 & 21 & 23 & 55 qubits \\
		2291 & 12 & 23 & 25 & 60 qubits \\
		10057 & 14 & 27 & 29 & 70 qubits 
	\end{tabular}	
\end{table}

\subsection{KratosVM Results}
The KratosVM is a register based quantum virtual machine that implements quantum instructions based on the multiplex network methods of QuMvNs. We used the bytecode specification outlined by Sakkaris \cite{sakkaris2018} which is capable of implementing Shor’s Algorithm. The KratosVM implements quantum arithmetic operations such as addition \(iquadd\), modular addition \(iquadd\_mod\) and controlled modular multiplication \(ciqumul\_mod\) by following the methods of Beckman et. Al. \cite{beckman1996efficient}, Vedral et. Al.\cite{vedral1996quantum} and Van Meter et. Al.\cite{van2005fast}. With this set of quantum arithmetic instructions, along with the rest of the instruction set specified in \cite{sakkaris2018}, we are able to implement Shor’s Algorithm as outlined in FIG.~\ref{fig:shor1} using the KratosVM. The bytecode program that implements the factorization of $N=10,057$ with random seed $a=4983$ is included in Appendix A.1 as an example. The bytecode implementation for all other simulation can be found on GitHub \cite{sakkaris2019Kratos01}.

Measurement on the KratosVM is treated as a sampling problem. We specify how many measurements we want to make and the KratosVM performs the measurements and gives results in terms of a frequency distribution. We do not output probability amplitudes or state vectors, only results of measurements which is a bit string along with a frequency of occurrence. This is very similar to how quantum hardware is designed to give measurement results. Throughout the computation and measurement, single-precision complex numbers are used for edge weights and double-precision complex numbers are used for the probability amplitudes of the layers. The efficiency of our simulation does not depend on the semi-prime number being factored, or the random seed chosen. We keep our methods as generic as possible so that the KratosVM can be extended to simulated a variety of quantum circuits, not just Shor’s Algorithm.

After measurement sampling was complete, we passed the results through the classical part of Shor’s Algorithm discussed in Section 4.1. In Shor’s Algorithm, every specific output value of the period finding algorithm has a probability of factoring the input or not factoring by finding an odd period or a common factor of $1$. We summarize the number of values sampled and how many values successfully factored the input in Table \ref{tab:table1}. We plotted all results in Figures \ref{fig:shor35} - \ref{fig:shor70}. A green point represents a measurement that factored the input $N$ and a red point a measurement that failed to factor the input $N$. The horizontal axis of the Figures represents the decimal integer value of the bit string measurement and the vertical axis represents the observed frequency of that bit string measurement.

\begin{table}[h!]
	\centering
	\caption{Runtime results: factorization of $N$}
	\label{tab:table1}
	\begin{tabular}{l|r|r|r|r|r|r|r|r} % <-- Alignments: 1st column left, 2nd middle and 3rd right, with vertical lines in between
		\textbf{qubits} & \textbf{$N$} & \textbf{seed $a$} & \textbf{samples} & \textbf{successes} & \textbf{\% success} & \textbf{time (sec)} & $p,q$ \\
		\hline
		35 & 77 & 69 & 500K & 246671 & 49.3\% & 0.267 & 7, 11\\
		40 & 145 & 73 & 500K & 244725 & 48.9\% & 0.500 & 5, 29\\
		50 & 731 & 426  & 500K & 247683 & 49.5\% & 17.000 & 17, 43\\
		55 & 1273 & 1229  & 500K & 214035 & 42.8\% & 72.000 & 67, 19\\
		60 & 2291 & 1301  & 500K & 233830 & 46.8\% & 310.000 & 79, 29 \\
		70 & 10057 & 4983  & 500K & 418253 & 83.7\% & 4740.000 & 89, 113
	\end{tabular}
\end{table}

\begin{figure}[h!]
	\caption[]{\textbf{Runtime (Sec) of Shor's Algorithm Simulation}}
	\includegraphics[width=\linewidth]{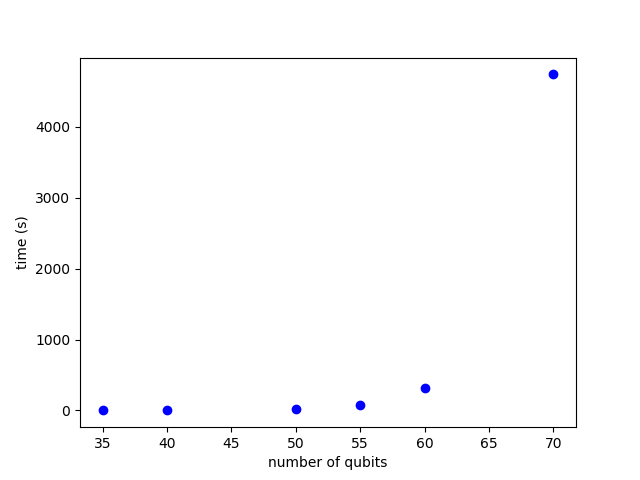}
	\label{fig:shorRuntime}
\end{figure}

\begin{figure}[h!]
	\caption[]{\textbf{Factorization of $N=77$ with $a=69$ using $35$ qubits}}
	\includegraphics[width=0.8\linewidth]{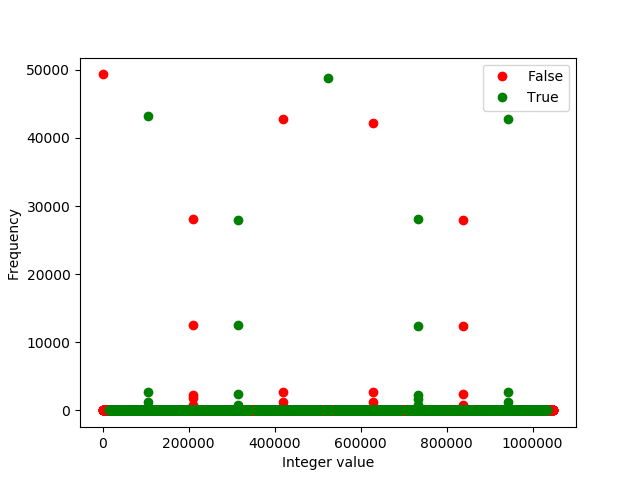}
	\label{fig:shor35}
\end{figure}

\begin{figure}[h!]
	\caption{\textbf{Factorization of $N=145$ with $a=73$ using $40$ qubits}}
	\includegraphics[width=0.8\linewidth]{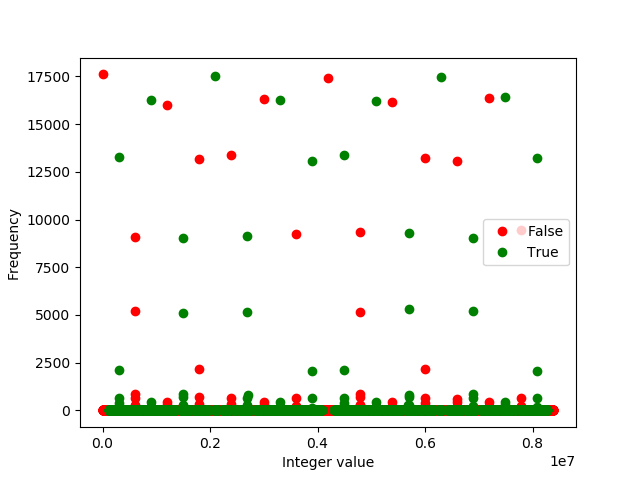}
	\label{fig:shor40}
\end{figure}

\begin{figure}[h!]
	\caption{\textbf{Factorization of $N=731$ with $a=426$ using $50$ qubits}}
	\includegraphics[width=0.8\linewidth]{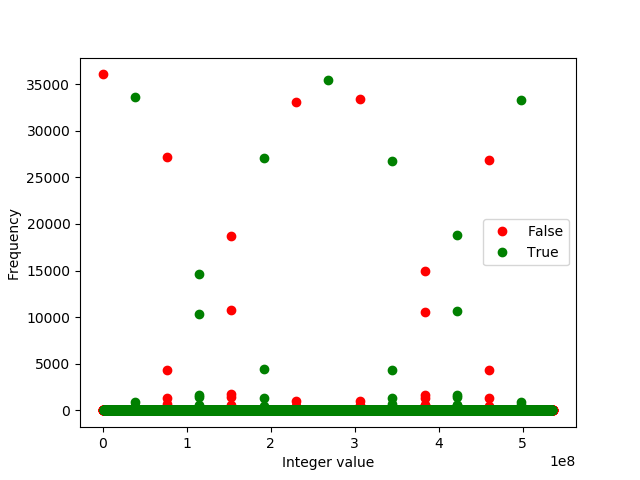}
	\label{fig:shor50}
\end{figure}

\begin{figure}[h!]
	\caption{\textbf{Factorization of $N=1,273$ with $a=1,229$ using $55$ qubits}}
	\includegraphics[width=0.8\linewidth]{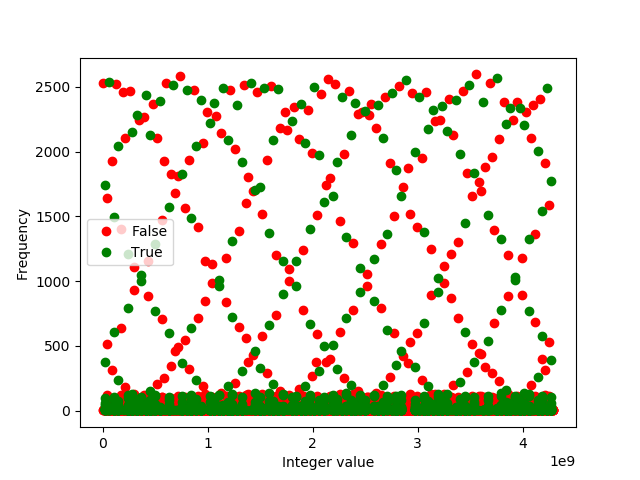}
	\label{fig:shor55}
\end{figure}

\begin{figure}[h!]
	\caption{\textbf{Factorization of $N=2,291$ with $a=1301$ using $60$ qubits}}
	\includegraphics[width=0.8\linewidth]{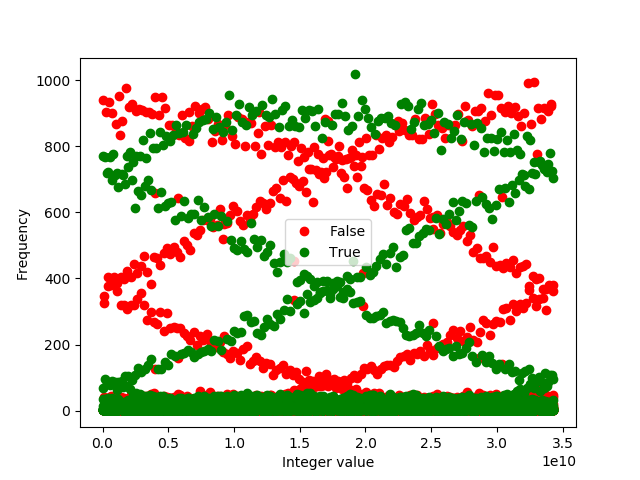}
	\label{fig:shor60}
\end{figure}

\begin{figure}[h!]
	\caption{\textbf{Factorization of $N=10,057$ with $a=4983$ using $70$ qubits}}
	\includegraphics[width=0.8\linewidth]{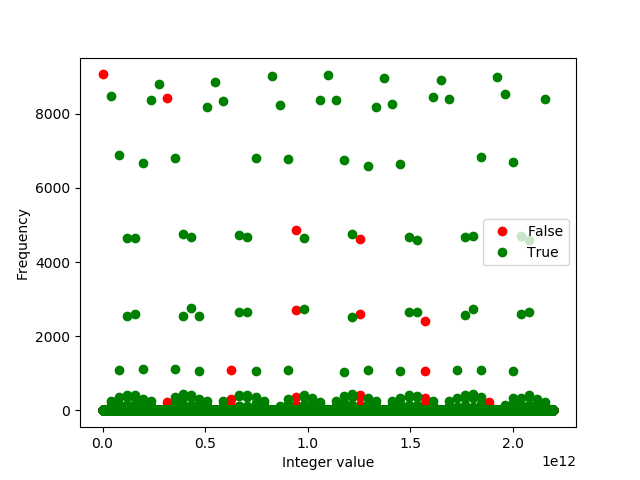}
	\label{fig:shor70}
\end{figure}

The $35$-$60$ qubit simulations were performed on a 2018 System76 Oryx Pro with an Intel Core i7-8750H CPU (6 cores, 12 threads) and 32GB of DDR4-2666 RAM. The $70$ qubit simulation was performed on a gcloud n1-highmem-96 (96 vCPUs, 624GB RAM) with an Intel Xeon Skylake CPU Platform. The fact that we can simulate a $60$-qubit instance of Shor's Algorithm on a portable workstation such as the Oryx Pro shows that the QuMvN formulation has significant advantages over traditional quantum simulation methods with respect to Shor's Algorithm. The classical simulation of Shor's Algorithm is still exponentially difficult (as seen in FIG. \ref{fig:shorRuntime}), however, QuMvNs allow us to push the boundary of feasibility regarding classical simulation of Shor's Algorithm. Also, the $70$-qubit simulation of Shor's Algorithm is the largest known simulation we are aware of and we did not require a supercomputer to perform the simulation.  

\clearpage
\captionsetup{justification=raggedright,singlelinecheck=false}

\hrulefill

\hrulefill

\hrulefill

	\clearpage
	\section{Conclusion and Future Work}
	
		We showed that the language of multiplex networks and information theory can be combined to describe wave functions for quantum computation and define it as a Quantum Multiverse Network (QuMvN). Each layer of a QuMvN is a discrete information source and the sum of layers is what Shannon called a ``mixed'' information source. We explained the parallels between layers of a QuMvN and worlds in the Many Worlds Interpretation. Thus, QuMvNs give us an information theoretic point of view of quantum mechanics. Measurements can be performed by sampling via random walks and we saw quantum logic gates can be defined as edge transformation of a QuMvNs. 
		 
		With the single qubit gates $H$, $R(k)$, $X$ and control-$U$ gates defined we have a universal gate set. In principle, we can simulate any quantum circuit using the QuMvN formalism, but why use the QuMvN formulation? As Feynman said with the path integral formulation of quantum mechanics: ``there is a pleasure in recognizing old things from a new point of view. Also, there are problems for which the new point of view offers a distinct advantage'' \cite{feynman2005space}. Problems for which QuMvNs offer a distinct advantage is in the classical simulation and analysis of quantum circuits, specifically, Shor's Algorithm.
		
		We used the QuMvN formalism to develop a classical simulation of Shor's Algorithm. We showed results of multiple classical simulations of Shor's Algorihtm successfully factoring the semi-prime numbers $77$, $145$, $731$, $1273$, $2291$ and $10,057$ using $35$, $40$, $50$, $55$, $60$ and $70$ qubits respectively. Simulation of $35$-$60$ qubits were performed on a portable workstation and the $70$-qubit simulation on a gcloud n1-highmem-96. Even though the classical simulation of Shor's Algorithm is exponentially difficult, the QuMvN formalism allows us to explore Shor's Algorithm with more qubits on impressively frugal hardware. Having the ability to run a $60$-qubit instance of Shor's Algorithm on a portable workstation with 32GB of RAM allows future quantum software engineers to experiment with this algorithm on a local development environment without the need of expensive quantum hardware. Also, larger instances of $70$-qubit simulation can be further explored on commodity cloud servers. We think classical simulation of Shor's Algorithm on this scale will be beneficial to quantum computing researchers and practitioners. Our future work will be focused on extending QuMvNs to other Quantum Algorithms with the same type of scale and success.
				
		If the number of layers in a QuMvN can be kept polynomial in the number of qubits $n$ and each layer has a unique set of paths then there is no quantum speedup and we can efficiently simulate with classical resources. Also, we explored problems in each level of the layer interaction hierarchy. Determining which problems fall into this which level of the hierarchy is the next challenge for QuMvNs. A fascinating question for further research is whether layers can be re-arranged or aggregated so that problems can move from one level of the hierarchy to another.
		
		Of particular interest will be the role of randomization algorithms in reducing the number of layers in a QuMvN. For example, methods of layer aggregation are useful in multiplex networks and may play a role in QuMvNs \cite{de2015structural}. Also, by relating layers in a QuMvN with Worlds in The Many Worlds Interpretation, we may be able to define the complexity of a wave function with the number of Worlds needed to represent it. Lastly, after running a $70$ qubit instance of Shor's Algorithm on a modest commodity cloud server, we will next explore RQC \cite{villalonga2018flexible} circuits on larger machines and see if we can approach the $70$ qubit hardware equivalent for a fraction of the cost thus pushing the boundary for quantum supremacy \cite{boixo2018characterizing}.

		\clearpage
\appendix
\section{Appendices}
\subsection{Shor's Algorithm Bytecode}
\begin{verbatim}
// The following bytecode implements Shor's Factorization Algorithm
// for the number 10,057 with random number 4983. Using this template
// we can derive bytecodes for factoring other numbers as explained
// in Section 4. 41 qubits are declared by the programmer, 29 qubits (2*l + 1)
// are used by the VM for arithmetic. 500,000 samples are measured
.qudot qubits=41, ensemble=500000

.gate main: args=0, regs=9, qubit_regs=7
// control qubits start/end, upper register k
iload r1, 1
iload r2, 27
// modulo multiplication qubits start/end, lower register l
iload r3, 28
iload r4, 41

qload_seq q0, 1, 27
qloadr q1, r3
qloadr q2, r4

// initialize state
hon q0
xon q2

// number to factor
iload r5, 10057
// random number
iload r6, 4983

// setup loop variable
move r7, r2
iload r9, 0
// modular exponatiation
ModExp:
printr r7
brlez r7, doneModExp
modpow r8, r6, r9, r5
//printr r8
qloadr q3, r7
ciqumul_mod r8, r5, q1, q2, q3
decr r7
incr r9
br ModExp

doneModExp:
// measure the second register
qload_seq q4, 28, 41
mon q4

printr r7
qloadr q5, r1
qloadr q6, r2
qft_inv q5, q6

printr r7
halt


\end{verbatim}
\end{document}